\documentclass[twocolumn,reprint,floatfix,amsmath,amssymb,twocolumn,english,aps,prl,10pt,superscriptaddress]{revtex4-2}
\usepackage[T1]{fontenc}
\usepackage{amsmath,graphicx,amssymb,epsfig,babel,dsfont}
\usepackage{mathrsfs}
\usepackage{color}
\usepackage{amsbsy}
\usepackage{float}

\def\bbm[#1]{\mbox{\boldmath $#1$}}

\begin{document}

\title{Control of the local photonic density of states above magneto-optical metamaterials}

\author{Philippe Ben-Abdallah}
\email{pba@institutoptique.fr}
\affiliation{Laboratoire Charles Fabry, UMR 8501, Institut d'Optique, CNRS, Universit\'{e} Paris-Saclay,
2 Avenue Augustin Fresnel, 91127 Palaiseau Cedex, France}

\date{\today}


\begin{abstract}
The local density of states  (LDOS) of electromagnetic field drives many basic processes associated to light-matter interaction such as the thermal emission of object, the spontaneous emission of quantum systems or the fluctuation-induced electromagnetic forces on molecules. Here, we study the LDOS in the close vicinity of magneto-optical metamaterials under the action of an external magnetic field and demonstrate that it can be efficiently changed in a narrow or a broad spectral range simply by changing the spatial orientation or the magnitude of this field. This result paves the way for an active control of the photonic density of states at deep-subwavelength scale.
\end{abstract}

\maketitle
\subsection*{I. INTRODUCTION}

Tailoring the photonic density of states  in the environnement of a textured solid at subwavelength scale is of prime importance to control the thermal emission of solids, the spontaneous emission and the decay rate (fluorescence) of quantum emitters placed in the neighborhood of these structures. This tuning can also be used to modify the force mediated by the vacuum fluctuations on neutral objects such as atoms or molecules.
Since the pioneer work of Purcell~\cite{Purcell} on the modification of the spontaneous emission of an object by changing its surrounding environnement many strategies have been proposed to sculpt the local density of states (LDOS) at a length scale smaller than the wavelength of electromagnetic field. Hence, nanophotonic structures such as photonic crystals~\cite{Yablonovitch,John}, surface gratings~\cite{Hesketh}, two dimensional systems~\cite{Koppens,Messina} or complex nanostructures~\cite{Viarbitskaya,Girard} were used to efficiently enhance or inhibit the spontaneous emission~\cite{Lodahl,Birowosuto} and the decay rates of molecules~\cite{Drexhage,Chance}.
In this paper we explore the possibility to tune the photonic states in the close vicinity of non-reciprocal metamaterials made with magneto-optical materials using an external magnetic field and show that the LDOS can be efficiently controled simply by changing the spatial orientation or the magnitude of this field.

\begin{figure}
\includegraphics[scale=0.28]{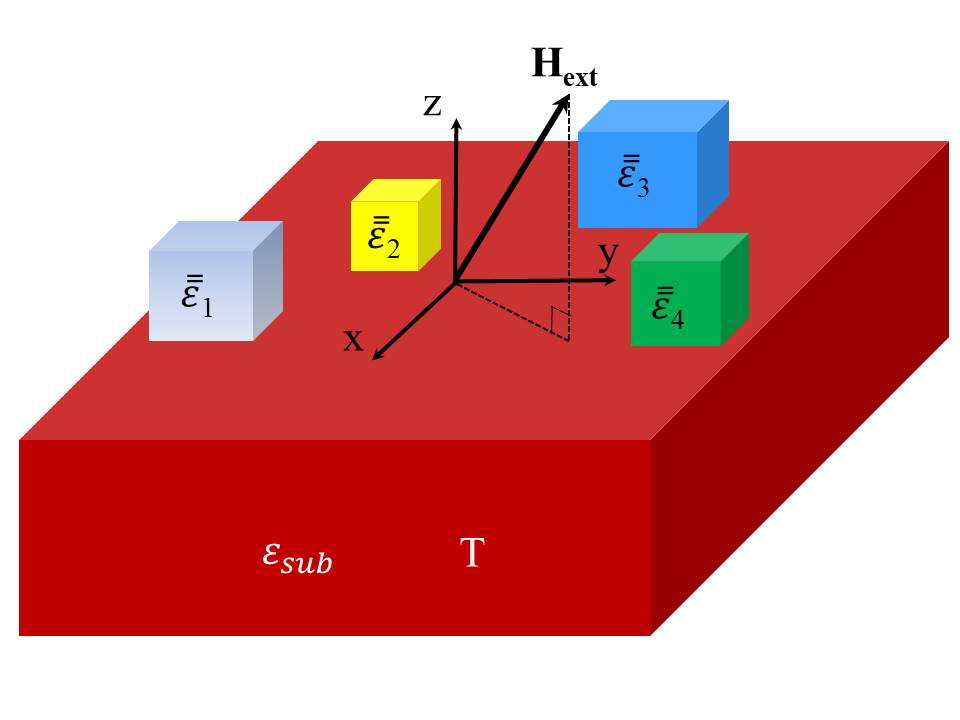}
\caption{ Sketch of a non-reciprocal metamaterial made with magneto-optical  nanostructures of permittivity $\bar{\bar{\boldsymbol{\varepsilon}}}_j$ deposited on an isotropic substrate of permittivity $\varepsilon_{sub}$. The structure is assumed at equilibrium at temperature $T$ with its surrounding environement and it is submitted to a uniform external magnetic field $\mathbf{H_{ext}}$ in an arbitrary angular orientation.}
\label{system}
\end{figure}

\subsection*{II. THEORY}
\subsubsection*{A. Geometry and optical properties}
To start, let us consider a non reciprocal metamaterial made with a network of magneto-optical nanostructures of permittivity $\bar{\bar{\boldsymbol{\varepsilon}}}_j$, deposited on an isotropic substrate of permittivivity $\varepsilon_{sub}$ as sketched on Fig.\ref{system}.
We  assume this system is at equilibrium at temperature $T$ with its surrounding environment and it is uniformly shined  by an external magnetic field $\bold{H_{ext}}$ in an arbitrary direction. When $\bold{H_{ext}}$ is parallel to the normal $\bold{z}$  of the subrate, the permittivity tensor associated to each magneto-optical nanostructure takes the following form in the canonical basis~\cite{Moncada}
\begin{equation}
\bar{\bar{\boldsymbol{\varepsilon}}}_j(H_{ext})=\left(\begin{array}{ccc}
\varepsilon_{j1}& -i\varepsilon_{j2} & 0\\
i\varepsilon_{j2} & \varepsilon_{j1} & 0\\
0 &  0 & \varepsilon_{j3}\\
\end{array}\right),\label{Eq:permittivity}
\end{equation}
where $\varepsilon_{j1}$, $\varepsilon_{j2}$ and $\varepsilon_{j3}$ generally depend on the magnitude $H_{ext}$ of magnetic field.
When the spatial orientation of $\bold{H_{ext}}$ is changed by rotations in the canonical basis $(\bold{x},\bold{y},\bold{z})$ as illustrated in Fig.1, the permittivity tensor associated to each nanostructure is readily derived from expression (\ref{Eq:permittivity}) by simple composition rules with the corresponding rotation matrices.
In the case  where $\bold{H_{ext}}$ is in the $(x,z)$ plane making an angle $\theta$ with the normal $\bold{z}$  to the surface the permittivity tensor of  magneto-optical material reads
\begin{equation}
\bar{\bar{\boldsymbol{\varepsilon}}}_j=
\left(\begin{array}{ccc}
\varepsilon_{j1}\cos^2\theta+\varepsilon_{j3}\sin^2\theta& -i\varepsilon_{j2}\cos\theta &\frac{1}{2}\sin(2\theta)(\varepsilon_{j1}-\varepsilon_{j3})\\
i\varepsilon_{j2}\cos\theta & \varepsilon_{j1} & i\varepsilon_{j2}\sin\theta\\
\frac{1}{2}\sin(2\theta)(\varepsilon_{j1}-\varepsilon_{j3}) & -i\varepsilon_{j2}\sin\theta  &\varepsilon_{j1}\sin^2\theta+\varepsilon_{j3}\cos^2\theta\\
\end{array}\right).\label{Eq:permittivity}
\end{equation}

\subsubsection*{B. Local density of states of electromagnetic field}
The LDOS $\rho(\bold{r},\omega)$ of electromagnetic field at a given point $\bold{r}=(x,y,z)$ and frequency $\omega$ can be calculated from the density of energy of electromagnetic field
\begin{equation}
u(\bold{r},\omega)=\epsilon_0<\mid \bold{E}(\bold{r},\omega)\mid^2>+\mu_0<\mid \bold{H}(\bold{r},\omega)\mid^2>, \label{energy_density}
\end{equation}
associated to the average value of local electric field $\bold{E}$ and magnetic field $\bold{H}$ using the general relation
\begin{equation}
u(\bold{r},\omega)=\rho(\bold{r},\omega)\Theta(\omega,T), \label{ldos_vs_energy}
\end{equation}
between this density and the LDOS. Here, $\Theta(T,\omega)={\hbar\omega}/[{e^{\frac{\hbar\omega}{k_B T}}-1}]$ denotes the mean energy of a harmonic oscillator at temperature $T$. When the  size of nanostructures is much smaller than the thermal wavelength  $\lambda_{th}=\hbar c/k_B T$ and the separation distance between these nanostructures is sufficiently large they behaves as  simple dipoles~\cite{pbaPRB2008,Rubi,Becerril}. In this case, the electric field can be related to the dipolar moments $\bold{p_i}$ of nanostructures as followed
\begin{equation}
  \mathbf{E}(\bold{r})=\omega^{2}\mu_{0}\underset{i=1}{\overset{N}{\sum}}\mathscr{G}^{EE}(\bold{r},\bold{r_i})\mathbf{p}_{i},
  \label{Eq:field}
\end{equation}
 with $\mu_{0}$ the vacuum permeability and $\mathscr{G}^{EE}(\bold{r},\bold{r'})$ is
the dyadic Green tensor between the point $\bold{r}$ and $\bold{r'}$
inside the set of N nanostructures. On the other hand each dipolar
moment can be decomposed into the form 
\begin{equation}
\mathbf{p}_{i}=\mathbf{p}_{i}^{fluc}+\mathbf{p}_{i}^{ind}\label{polar}
\end{equation}
where the first term of rhs is its fluctuating part while the second term is the part induced by all others dipolar and it reads
\begin{equation}
  \mathbf{p}_{i}^{ind}=k_0^2\bar{\bar{\boldsymbol{\alpha}}}_{i}\underset{j\neq i}{\sum}\mathscr{G}^{EE}(\bold{r_i},\bold{r_j})\mathbf{p}_j\label{Eq:Polarizability}
\end{equation}
 $\bar{\bar{\boldsymbol{\alpha}}}_{i}$ being the nanostructure polarizability, $k_0=\frac{\omega}{c}$ the wavenumber in vacuum and $\varepsilon_{0}$
is the vacuum permittivity. 
Inserting expression (\ref{Eq:Polarizability}) into relation (\ref{polar}) it immediately follows the relation beween de dipolar moments and their fluctuating part
\begin{equation}
\mathbf{p}_i=\underset{j=1}{\overset{N}{\sum}}\mathds{T}^{-1}_{EE,ij}\mathbf{p}_j^{fluc}
\label{dipole_vs_fluc}
\end{equation}
where $\mathds{T}_{EE}$ is a block matrix of component
\begin{equation}
\mathds{T}_{EE,ij} =\delta_{ij}\mathds{1}-(1-\delta_{ij})k_0^2 \bar{\bar{\boldsymbol{\alpha}}}_{i} \mathscr{G}^{EE}(\bold{r_i},\bold{r_j}).
\label{Eq:A0}
\end{equation}
It follows that the local electric fied can be written in term of fluctuating dipolar moments as~\cite{pbaPRL2011,pbaRMP}
\begin{equation}
  \mathbf{E}(\bold{r})=\omega^{2}\mu_{0}\underset{j=1}{\overset{N}{\sum}}\mathds{G}^{EE}(\bold{r},\bold{r_j})\mathbf{p}_{j}^{fluc},
  \label{Eq:field_fluc}
\end{equation}
where the full electric-electric Green tensor takes the form
\begin{equation}
  \mathds{G}^{EE}(\bold{r},\bold{r_j})=\underset{i=1}{\overset{N}{\sum}}\mathscr{G}^{EE}(\bold{r},\bold{r_i})\mathds{T} _{EE,ij}^{-1}.
  \label{full_Green_electric}
\end{equation}
For an ensemble of dipoles  in free space the propagator reads
\begin{equation}
\begin{split}
\mathscr{G}^{EE}(\bold{r'},\bold{r''})\equiv\mathscr{G}_0^{EE}(\bold{r'},\bold{r''})=\:\:\:\:\:\:\:\:\:\:\:\:\:\:\:\:\:\:\:\:\:\:\:\:\:\:\:\:\:\:\:\:\:\:\:\:\:\:\:\:\:\:\:\:\:\:\:\:\:\:\:\:\:\:\:\:\:\\
\frac{\exp(ik_0 r)}{4\pi r}\left[\left(1+\frac{ik_0 r-1}{k_0^{2}r^{2}}\right)\mathds{1}+\frac{3-3ik_0 r-k_0^{2}r^{2}}{k_0^{2}r^{2}}\widehat{\mathbf{r}}\otimes\widehat{\mathbf{r}}\right],
\label{propagator_vacuum}
\end{split}
\end{equation}
where $\widehat{\mathbf{r}}\equiv\mathbf{\bold{r}}/r$,
$\mathbf{r}=\bold{r'}-\bold{r''}$ and  $r=\mid\mathbf{r}\mid$ and $\mathds{1}$
stands for the unit dyadic tensor. 
When the dipoles are located in vacuum above a solid material the propagator reads
\begin{equation}
\mathscr{G}^{EE}(\bold{r'},\bold{r''})\equiv\mathscr{G}_0^{EE}(\bold{r'},\bold{r''})+\mathscr{G}^{EE,sc}(\bold{r'},\bold{r''}),\label{propagator_interface}
\end{equation}
where the second term of rhs describe the scattering by the interface between the vacuum and material. 

\subsubsection*{C. Electric-electric dyadic Green tensor for the reflected field}

\begin{figure}
\includegraphics[scale=0.28]{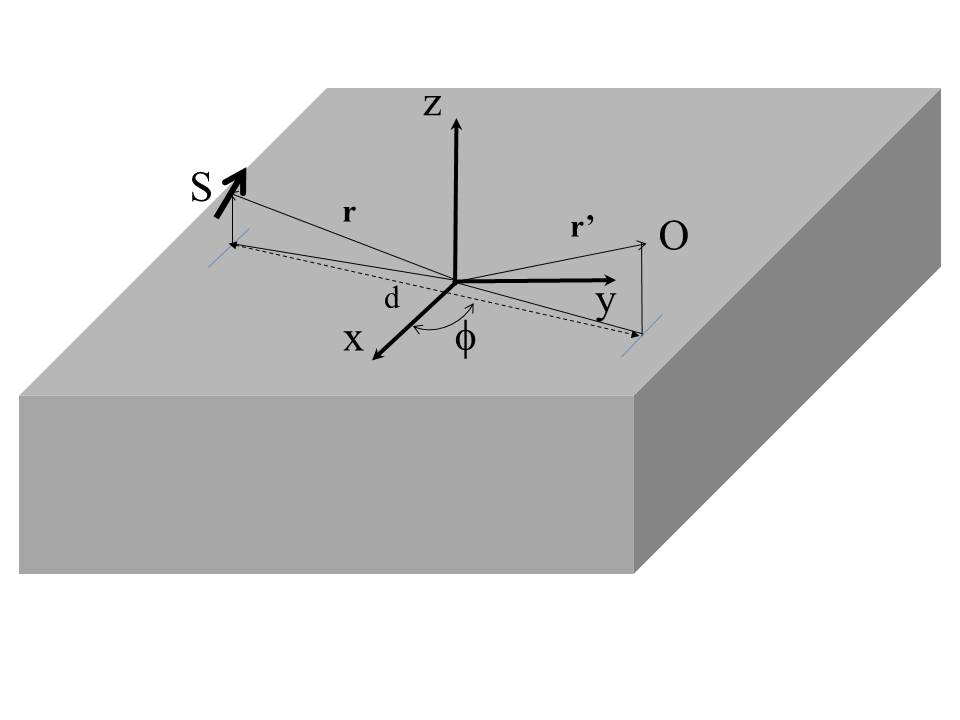}
\caption{ Coordinates system associated to the Green tensor between a  source $S$ (dipole)  located at point $\bold{r}$ and an obseravtion point $O$ at $\bold{r'}$ above a substrate.}
\label{coordinate}
\end{figure}

When the dipoles are located in vacuum above a solid material the propagator reads
\begin{equation}
\mathscr{G}^{EE}(\bold{r'},\bold{r''})\equiv\mathscr{G}_0^{EE}(\bold{r'},\bold{r''})+\mathscr{G}^{EE,sc}(\bold{r'},\bold{r''}),\label{propagator_interface}
\end{equation}
where the second term of rhs describe the scattering by the interface between the vacuum and material. 
This scattering term can be written as an integral with respect to the modulus $\kappa$ of the modulus of the parallel component $\bold{\kappa}=(kx, ky)$ of wavevector (i.e. parallel to the  $x- y$ plane) (see Fig.\ref{coordinate}) as~\cite{Novotny}
\begin{align}
 \mathscr{G}^{EE,sc}(\mathbf{r'},\bold{r''})\\
=\int_{0}^{\infty}\frac{d\kappa}{2\pi}\kappa\frac{i}{2k_z}exp(ik_z\mid z+z'\mid)[r^s\mathds{S}_{EE}+r^p\mathds{P}_{EE}]\label{scat}
\end{align}
where $k_z =\sqrt{k_0^2-\kappa^2}$ is the normal component of the wavevector in vacuum while $r^s=\frac{k_z-k_{z2}}{k_z+k_{z2}}$ and $r^p=\frac{\varepsilon_2k_z-k_{z2}}{\varepsilon_2 k_z+k_{z2}}$ are the ordinary Fresnel coefficients of the interface associated to waves of polarization $s$ and $p$, respectively, $k_{z2}=\sqrt{\varepsilon_2 k_0^2-\kappa^2}$ being the normal component of wavector inside the substrate. 
In this expression the scattering matrix $\mathds{S}$ and $\mathds{P}$ can be readily calculated using the Sipe formalism~\cite{Sipe} and they read
\begin{equation}
\mathds{S}_{EE}=\mathscr{\left(\begin{array}{ccc}
A & -C & 0\\
-C & B & 0\\
0& 0 & 0\\
\end{array}\right)},
\label{matrix-S_EE}
\end{equation}
\begin{equation}
\mathds{P}_{EE}=\frac{1}{k_0^2}\mathscr{\left(\begin{array}{ccc}
  -k_z^2 B &  -k_x^2 C &  -k_z\kappa E\\ 
 -k_x^2 C &  -k_z^2 A & -k_z\kappa D\\ 
k_z\kappa E & k_z\kappa D & \kappa^2 F\\
\end{array}\right)},
\label{matrix-P_EE}
\end{equation}
with
\begin{equation}
A=\frac{1}{2}[J_0(\kappa d)+J_2(\kappa d)\cos(2\phi)],\label{A}
\end{equation}
\begin{equation}
B=\frac{1}{2}[J_0(\kappa d)-J_2(\kappa d)\cos(2\phi)],\label{B}
\end{equation}
\begin{equation}
C=\frac{1}{2}J_2(\kappa d)\sin(2\phi),\label{C}
\end{equation}
\begin{equation}
D=i J_1(\kappa d)\sin(\phi),\label{D}
\end{equation}
\begin{equation}
E=i J_1(\kappa d)\cos(\phi),\label{E}
\end{equation}
\begin{equation}
F=J_0(\kappa d),\label{F}
\end{equation}
$J_i$ denoting the Bessel function of order $i$, $d=\sqrt{(x'-x'')^2+(y'-y'')^2}$ is the distance between the projection of two points $\bold{r'}$ and $\bold{r''}$ on the $x- y$ plane and $\phi$ is the angle of this projected vector and the $x$-axis (Fig.\ref{coordinate}).

\subsubsection*{D. Magnetic-electric dyadic Green tensor for the reflected field}
Let us start by considering the case of a set of dipoles in vacuum. The local  magnetic field $\mathbf{H}$ is related to the electric field $ \mathbf{E}$ by the Faraday relation
\begin{equation}
\mathbf{H}=\frac{i}{\omega\mu_0}\nabla\times\mathbf{E}.\label{H vs E}
\end{equation}
With a single electric dipole the electric field reads
\begin{equation}
\mathbf{E}=\omega^2\mu_0\mathscr{G}_0^{EE}\mathbf{p}\label{E vs p}
\end{equation}
so that
\begin{equation}
\mathbf{H}=i\omega\nabla\times\mathscr{G}_0^{EE}\mathbf{p}\label{E vs p}.
\end{equation}
Thus it immediately follows 
\begin{equation}
\mathscr{G}_0^{HE}=\nabla\times\mathscr{G}_0^{EE}.\label{G_HE vs G_EE}
\end{equation}
After a straightforward calculation we get
\begin{equation}
\mathscr{G}_0^{HE}(\bold{r'},\bold{r''})=\frac{\exp(ik_0 r)}{4\pi r}k_0\frac{\widehat{\mathbf{r}}\otimes\mathds{1}}{r}(i-\frac{1}{k_0 r}),\label{propagator_vacuum}
\end{equation}
where $\mathbf{r}\otimes\mathds{1}$ denotes the cross product of $\mathbf{r}$ with each column of the unit dyadic tensor.
Following the same reasoning as for the electric-electric Green tensor  the full magnetic-electric Green tensor reads
\begin{equation}
  \mathds{G}^{HE}(\bold{r},\bold{r_j})=\underset{i=1}{\overset{N}{\sum}}\mathscr{G}^{HE}(\bold{r},\bold{r_i})\mathds{T} _{HE,ij}^{-1},
  \label{full_Green_magnetic}
\end{equation}
where $\mathds{T} _{HE}$ is obtained by substituting $\mathscr{G}^{EE}\rightarrow\mathscr{G}^{HE}$ into $\mathds{T} _{EE}$. It follows that the local magnetic fields reads in terms of fluctuating electric dipoles
\begin{equation}
  \mathbf{H}(\bold{r})=i\omega\underset{j=1}{\overset{N}{\sum}}\mathds{G}^{HE}(\bold{r},\bold{r_j})\mathbf{p}_{j}^{fluc}.
  \label{Eq:mag_fluc}
\end{equation}
Above an interface we must add the scattering part of Green tensor to the propagator in vaccum which is given by
\begin{equation}
\begin{split}
 \mathscr{G}^{HE,sc}(\bold{r'},\bold{r''})= \nabla\times\mathscr{G}^{EE,sc}(\bold{r'},\bold{r''})\\
=\int_{0}^{\infty}\frac{d\kappa}{2\pi}\kappa\frac{i}{2k_z}exp(ik_z\mid z+z'\mid)[r^s\mathds{S}_{HE}+r^p\mathds{P}_{HE}].\label{scat}
\end{split}
\end{equation}
In this expression 
\begin{equation}
\mathds{S}_{HE}=\mathscr{\left(\begin{array}{ccc}
 i k_z C& -i k_z B & 0\\
i k_z A & -i k_z C & 0\\
0& 0 & 0\\
\end{array}\right)}
\label{matrix-S_HE}
\end{equation}
and
\begin{equation}
\mathds{P}_{HE}=\mathscr{\left(\begin{array}{ccc}
 i k_z^3 C &  i k_z^3 A & i k_z^2\kappa D\\ 
-i k_z^3 B &  -i k_z^3 C & -i k_z^2\kappa E\\ 
0 & 0 & 0\\
\end{array}\right)}.
\label{matrix-P_HE}
\end{equation}

\subsubsection*{E. Correlations functions of fields}
It follows from expressions (\ref{Eq:field_fluc}) and  (\ref{Eq:mag_fluc}) that the correlation functions of electric and magnetic fields reads
\begin{equation}
\begin{split}
 <\mid \bold{E}(\bold{r},\omega)\mid^2>=\omega^4\mu_0^2\:\:\:\:\:\:\:\:\:\:\:\:\:\:\:\:\:\:\:\:\:\:\:\:\:\:\:\:\:\:\:\:\:\:\\
\times\underset{i,j}{\sum}\underset{l,k,n}{\sum}\mathds{G}^{EE}_{lk}(\bold{r},\bold{r_i})\mathds{G}^{EE*}_{ln}(\bold{r},\bold{r_j})<p^f_{i,k}p^{f*}_{j,n}>,
\label{correlation_Electric}
\end{split}
\end{equation}
\begin{equation}
\begin{split}
 <\mid \bold{H}(\bold{r},\omega)\mid^2>=-\omega^2\:\:\:\:\:\:\:\:\:\:\:\:\:\:\:\:\:\:\:\:\:\:\:\:\:\:\:\:\:\:\:\:\:\:\\
\times\underset{i,j}{\sum}\underset{l,k,n}{\sum}\mathds{G}^{HE}_{lk}(\bold{r},\bold{r_i})\mathds{G}^{HE*}_{ln}(\bold{r},\bold{r_j})<p^f_{i,k}p^{f*}_{j,n}>,
\label{correlation_Magnetic}
\end{split}
\end{equation}
where $\mathds{G}^{AB}_{lk}$ denotes the $lk$ component of Green tensor and $p^f_{i,k}$ is the $k^{th}$ component of $i^{th}$ fluctuating dipole. 
But according to the fluctuation dissipation theorem~\cite{Callen}, the correlations functions of dipolar moments read
\begin{equation}
\langle p^{f}_{i,l}p_{j,n}^{f*}\rangle=\frac{\epsilon_{0}}{i\omega}(\bar{\bar{\boldsymbol{\alpha}}}_{i,ln}-\bar{\bar{\boldsymbol{\alpha}}}^*_{i,nl})\Theta(\omega,T_{i})\delta_{ij}\delta_{ln},
\label{FDT}
\end{equation}
where $\Theta(T,\omega)={\hbar\omega}/[{e^{\frac{\hbar\omega}{k_B T}}-1}]$ denotes the mean energy of a harmonic oscillator at temperature $T$, $\bar{\bar{\boldsymbol{\alpha}}}_{i}$ is the polarizability associated to the $i^{th}$ dipole and $\delta_{\alpha\beta}$ is the usual Kronecker symbol.
As, the polarizability tensor is concerned,  it can be described, by taking into account the radiative corrections, using the following expression~\cite{Albaladejo}
\begin{equation}
\bar{\bar{\boldsymbol{\alpha}}}_{i}(\omega)=\left( \bar{\bar{\boldsymbol{1}}}-i\frac{k^3}{6\pi} \bar{\bar{\boldsymbol{\alpha_0}}}_{i}\right)^{-1} \bar{\bar{\boldsymbol{\alpha_0}}}_{i}\label{Eq:Polarizability2},
\end{equation}
where $ \bar{\bar{\boldsymbol{\alpha_0}}}_{i}$ denotes the quasistatic polarizability of the $i^{th}$ particle and $k=\omega/c$, $c$ being the speed of light in vacuum. For spherical particles in vacuum, the quasistatic polarizability takes the simple form
\begin{equation}
 \bar{\bar{\boldsymbol{\alpha_0}}}_{i}(\omega)=4\pi R^3\big(\bar{\bar{\boldsymbol{\varepsilon}}}_i-\bar{\bar{\boldsymbol{1}}}\big)\big(\bar{\bar{\boldsymbol{\varepsilon}}}_i+2\bar{\bar{\boldsymbol{1}}}\big)^{-1}\label{Eq:Polarizability2},
\end{equation}
where $R$ is the radius of particles. Of course others shapes can be considered without changing the general formalism previously introduced.

\subsubsection*{F. Resonance frequencies of a small magneto-optical particle}

Let us consider a simple particle of permittivity $\bar{\bar{\boldsymbol{\varepsilon}}}$ under the action of an external magnetic field $\bold{H_{ext}}$ which is in the $(x,z)$ plane and makes an angle $\theta$ with the $\bold{z}$  axis. The resonance frequencies of the particle are the frequencies where the polarizability diverges without loss.
\begin{figure}
\includegraphics[scale=0.32]{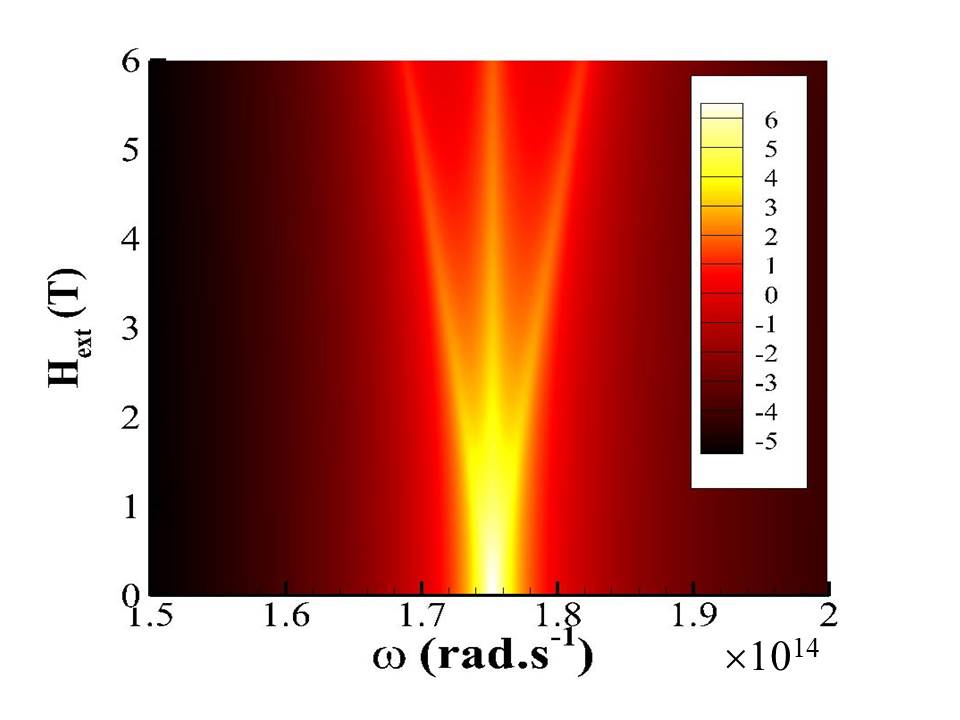}
\caption{ Resonances frequencies of an InSb particle in the $(\omega,H_{ext})$ plane. The plot shows the function $f(\omega,H_{ext})=\ln\left(|(\varepsilon_3+2)[(\varepsilon_1+2)^2-\varepsilon^2_2]|^{-1}\right)$.}
\label{dispersion-1p}
\end{figure}
But, according to expression (\ref{Eq:Polarizability2}), the quasi-static polarizability is resonant in vacuum when the matrix $\big(\bar{\bar{\boldsymbol{\varepsilon}}}+2\bar{\bar{\boldsymbol{1}}}\big)^{-1}$ diverges that is when $det\big(\bar{\bar{\boldsymbol{\varepsilon}}}+2\bar{\bar{\boldsymbol{1}}}\big)=0$. 
It is easy to show, after a straighforward calculation that
\begin{equation}
det\big(\bar{\bar{\boldsymbol{\varepsilon}}}+2\bar{\bar{\boldsymbol{1}}}\big)=adf+ae^2+b^2f-2bce-c^2d,\label{det}
\end{equation}
with
\begin{equation}
a=\varepsilon_{1}\cos^2\theta+\varepsilon_{3}\sin^2\theta+2,\label{a}
\end{equation}
\begin{equation}
b=i\varepsilon_{2}\cos\theta,\label{b}
\end{equation}
\begin{equation}
c=\frac{1}{2}\sin(2\theta)(\varepsilon_{1}-\varepsilon_{3}),\label{c}
\end{equation}
\begin{equation}
d=\varepsilon_{1}+2,\label{d}
\end{equation}
\begin{equation}
e=i\varepsilon_{2}\sin\theta,\label{e}
\end{equation}
\begin{equation}
f=\varepsilon_{1}\sin^2\theta+\varepsilon_{3}\cos^2\theta+2,\label{f}
\end{equation}
When $\theta=0$ (i.e. $\bold{H_{ext}}$ is paralel to $\bold{z}$) then $c=e=0$ and the resonant modes frequencies are the roots of the following equation
\begin{equation}
f(ad+b^2)=0.\label{reson}
\end{equation}
Written in term of components of  permittivy tensor, this equation decomposes into 
\begin{equation}
 \varepsilon_{3}(\omega)+2=0\label{Eq:res1}
\end{equation}
and
\begin{equation}
\left[ \varepsilon_{1}(\omega,H_{ext})+2\right]^2-\varepsilon_{2}^2(\omega,H_{ext})=0\label{Eq:res2}.
\end{equation}
The first equation  is independent on the magnetic field  (it corresponds to the vertical bright branch in Fig.\ref{dispersion-1p})  and it corresponds to the classical  resonance condition for a reciprocal spherical particle  of permittivity $\varepsilon_{3}$ in vacuum.  On the other hand, Eq.~\eqref{Eq:res2} can be decomposed into  two distincts equations
\begin{equation}
\varepsilon_{1}(\omega,H_{ext})-\varepsilon_{2}(\omega,H_{ext})+2=0,\label{Eq:res3}
\end{equation}
\begin{equation}
\varepsilon_{1}(\omega,H_{ext})+\varepsilon_{2}(\omega,H_{ext})+2=0,\label{Eq:res4}
\end{equation}
which both depend on the magnitude of external magnetic field as shown in Fig.\ref{dispersion-1p} for the two oblique branches. This spectral spliiting is similar to the Zeeman splitting of the electron levels inside a magnetic field. 
When $\theta=\frac{\pi}{2}$ (i.e. $\bold{H_{ext}}$ is paralel to $\bold{x}$) then $b=c=0$ and the resonant modes frequencies are the roots of the following equation
\begin{equation}
a(df+e^2)=0.\label{reson}
\end{equation}
which gives exactly the same roots as Eqs. (\ref{Eq:res3}) and (\ref{Eq:res4}) confirming so that the resonance frequencies of a single magneto-optical particle does not obviously depends on the orientation of external magnetic field. 
Of course, this is not true anymore in presence of other particles where these resonances closely depend on the interplay between the different particles and therefore on their spatial distribution.

\subsection*{III. TAILORING THE LDOS}
To show the potential of non-reciprocal metamaterials for controlling the photonic states we calculate the  LDOS in the neighborhood of simple networks. Here the goal is not to make an exhaustive study of the LDOS tailoring but to simply highlight the potential of magneto-optical metamaterials to control it. To this end we study networks  made with magneto-optical  nanoparticles of indium antimoniure (InSb) . 
In this case~\cite{Palik}
\begin{eqnarray}
\varepsilon_{1}(H_{ext})\!\!&=&\!\!\varepsilon_\infty\bigg(1+\frac{\omega_L^2-\omega_T^2}{\omega_T^2-\omega^2-i\Gamma\omega}+\frac{\omega_p^2(\omega+i\gamma)}{\omega[\omega_c^2-(\omega+i\gamma)^2]}\bigg) \label{Eq:permittivity1},\nonumber\\
\varepsilon_{2}(H_{ext})\!\! &=& \!\!\frac{\varepsilon_\infty\omega_p^2\omega_c}{\omega[(\omega+i\gamma)^2-\omega_c^2]}\label{Eq:permittivity2},\nonumber\\
\varepsilon_{3} \!\!&=&\!\! \varepsilon_\infty\bigg(1+\frac{\omega_L^2-\omega_T^2}{\omega_T^2-\omega^2-i\Gamma\omega}-\frac{\omega_p^2}{\omega(\omega+i\gamma)}\bigg)\label{Eq:permittivity3}.
\end{eqnarray}
For the numerical applications we assume that the dielectric constant at infinite-frequency  is $\varepsilon_\infty=15.7$ , the longitudinal optical phonon frequency$\omega_L=3.62\times10^{13}\,\mathrm{rad\,s}^{-1}$ is , the transverse optical phonon frequency is $\omega_T=3.39\times10^{13}\,\mathrm{rad\,s}^{-1}$, the plasma frequency of free carriers of density $\omega_p=(\frac{ne^2}{m^*\varepsilon_0\varepsilon_\infty})^{1/2}$ with the density  $n=1.36\times10^{19}\,\mathrm{cm}^{-3}$.
The effective mass is $m^*=7.29\times 10^{-32}\,\mathrm{kg}$, $\varepsilon_0$ being the vacuum permittivity, $\Gamma=5.65\times10^{11}\,\mathrm{rad\,s}^{-1}$ is the phonon damping constant,$\gamma=3.39\times10^{12}\,\mathrm{rad\,s}^{-1}$ is the free carrier damping constant, and $\omega_c=eH_{ext}/m^*$ is the cyclotron frequency with $e$ the electron charge. 

\subsubsection*{A. Single particle}

\begin{figure}[H]
\includegraphics[scale=0.3]{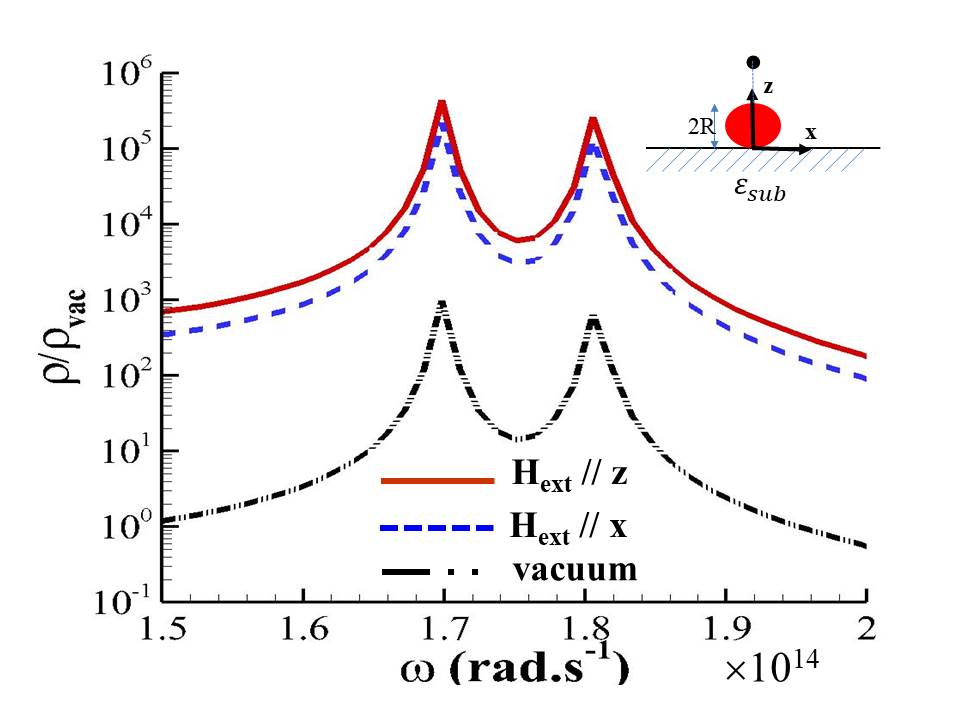}
\caption{ LDOS of electromagnetic field  at  $z=3R$ in vacuum above a InSb spherical particle of radius $R=50\:nm$  placed on a substrate  for an  external magnetic field $\mathbf{H_{ext}}$ (magnitude $H_{ext}=5T$) oriented along the $\mathbf{x}$ and $\mathbf{z}$ axis. The substrate is a semi-infinite transparent substrate ($\varepsilon_{sub}=4$). The double dashed curve corresponds to the LDOS above the InSb particle without substrate. The system is at equilibrium at temperature $T=300\,K$. The LDOS $\rho$  is normalized by the LDOS  $\rho_{vac}=\frac{\omega^2}{\pi^2 c^3}$ in vacuum.}
\label{ldos1particle}
\end{figure}

We first consider the simplest case shown in Fig.\ref{ldos1particle} where a single InSb nanoparticle is deposited on a transparent substrate ($\varepsilon_{sub}=4$) and we calculate the LDOS  above this particle  when the  external magnetic field  is rotated from the direction normal to the substrate surface to the direction parallel to it. 
We observe that the LDOS is mainly pronounced at the resonance frequencies solutions of Eq. (\ref{Eq:res3}) and (\ref{Eq:res4}) . Moreover the presence of a substrate enhances significantly the LDOS. This result  is well known since the pioneering works of Drexhage et al. \cite{Drexhage} and Chance et al. \cite{Chance} on the molecular fluorescence emission  of excited
atoms or molecules which strongly depends on their close environment. On the other hand, the  LDOS above the particle is qualitatively independent on the orientation of applied magnetic field (Fig.\ref{ldos1particle}).
Indeed, in a single particle system  there is no configurational resonance~\cite{KellerEtAl1993}, so that the LDOS spectrum keep the same shape when the magnetic field is rotated.
But, the situation radically changes in more complex networks where the interplay between the different particles (i.e. collective modes) start to play a role. 

\subsubsection*{B. Dimer of particles}

\begin{figure}[H]
\includegraphics[scale=0.3]{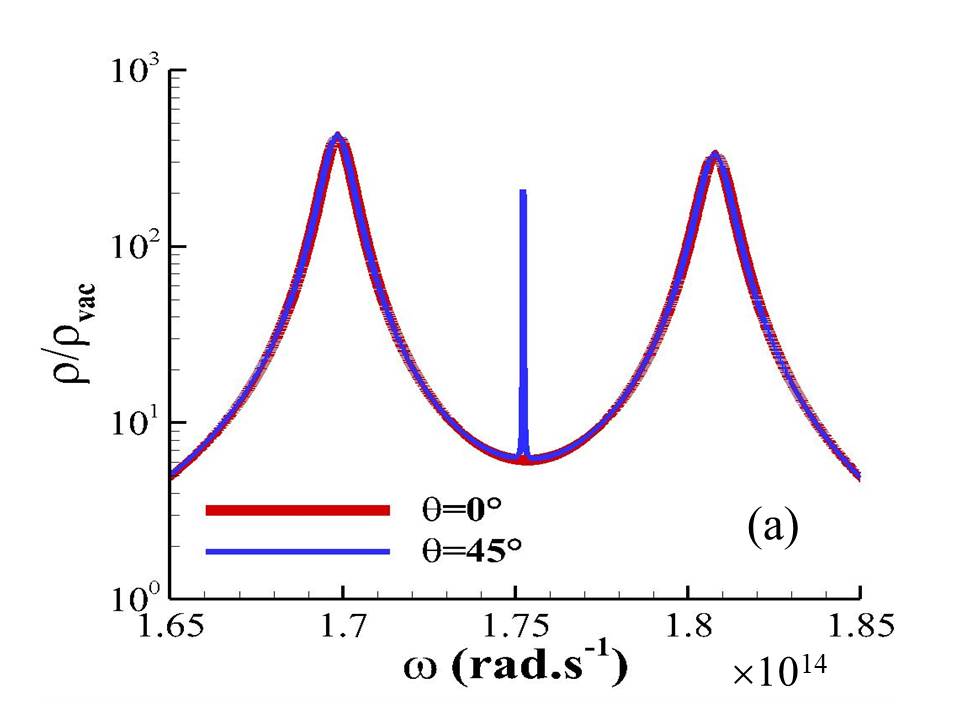}
\includegraphics[scale=0.3]{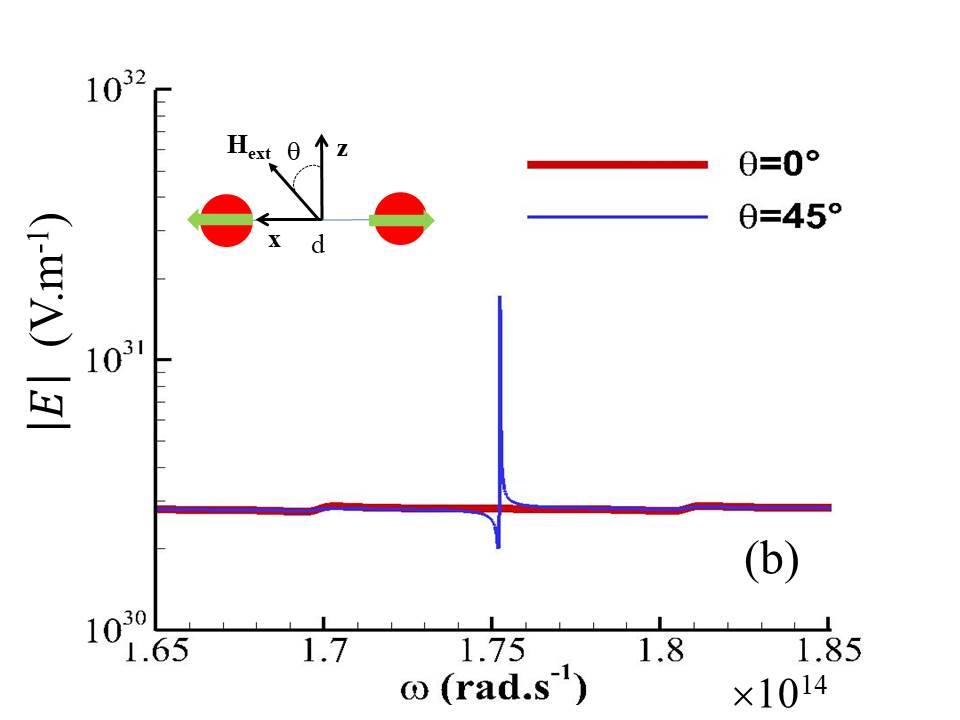}
\includegraphics[scale=0.3]{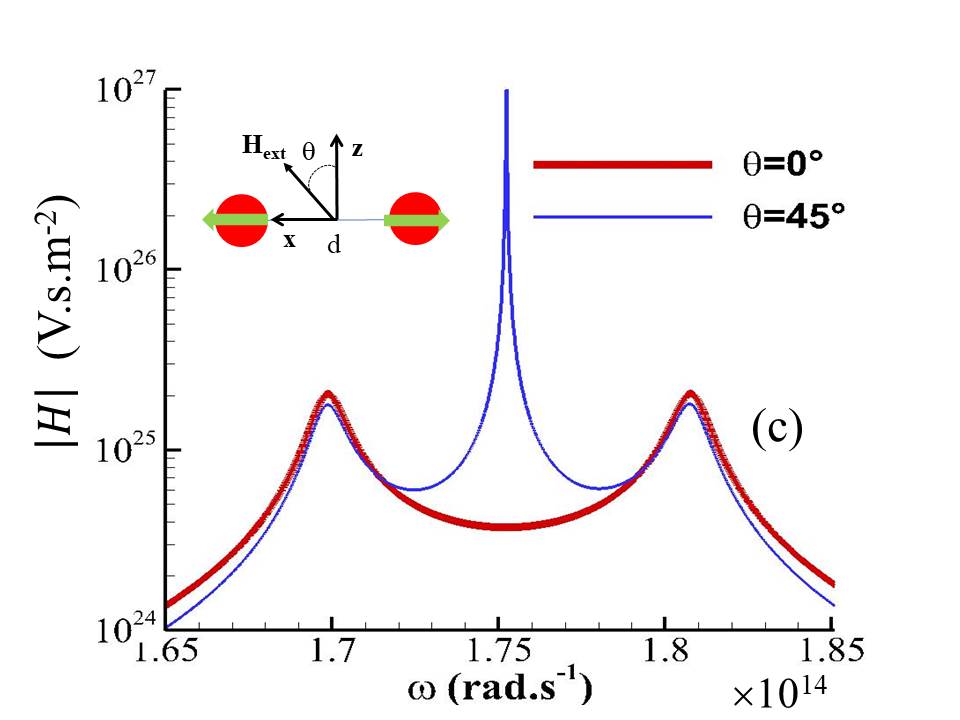}
\caption{(a)  LDOS at $z=3R$ of electromagnetic field above  a dimer of InSb particles (radius $R=50\:nm$)  in vacuum separated by a distance $d=6R$ under the action of an external magnetic field $H_{ext}$ making an angle $\theta$ with the $\bold{z}$ axis in the ($\bold{x}$,$\bold{z}$) plane. (b)  Magnitude of electric field  radiated between the two particles at $z=3R$ by two permanent (nomalized) dipoles oriented in opposite direction along the $\bold{x}$ axis . (c) Magnitude of magnetic field in the same conditions.  The magnitude of applied magnetic field is $H_{ext}=5 T$.}
\label{ldos-dimer-vac}
\end{figure}

\begin{figure}
\includegraphics[scale=0.3]{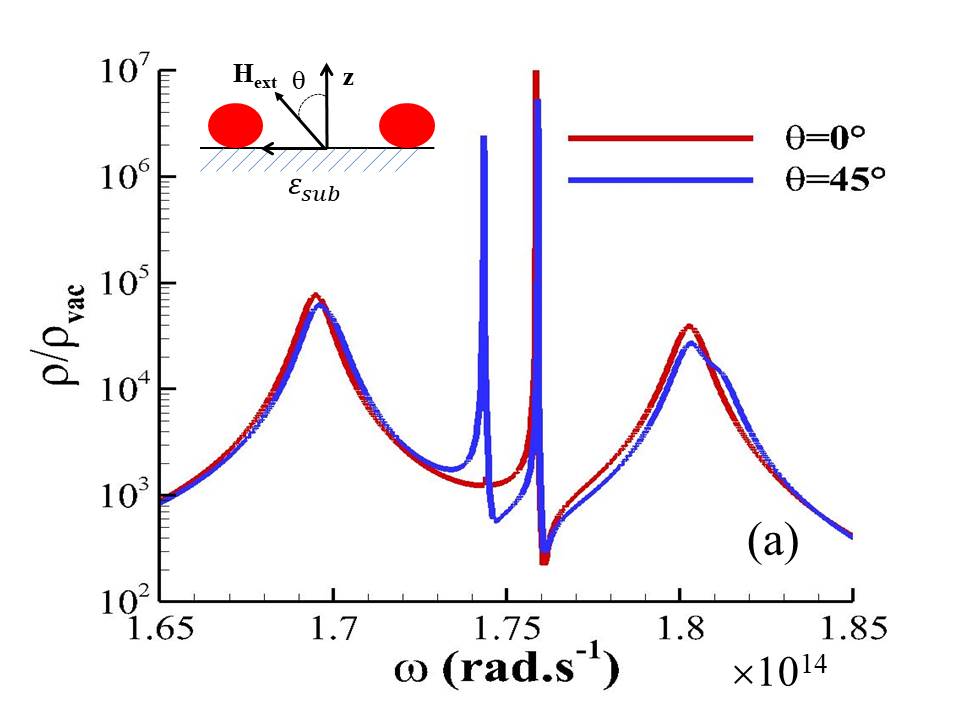}
\includegraphics[scale=0.3]{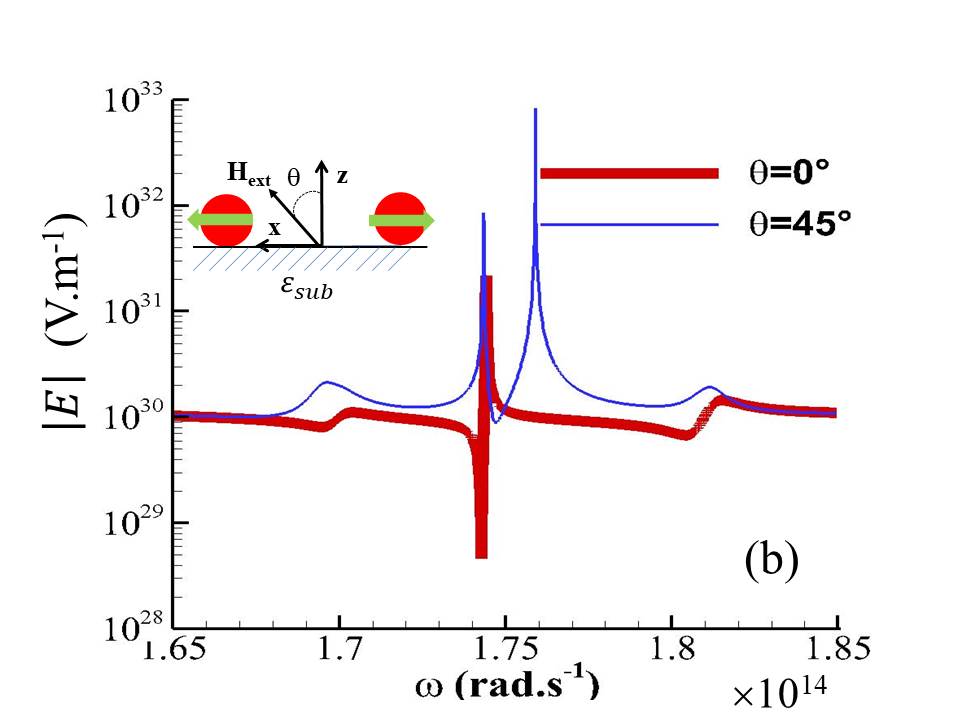}
\includegraphics[scale=0.3]{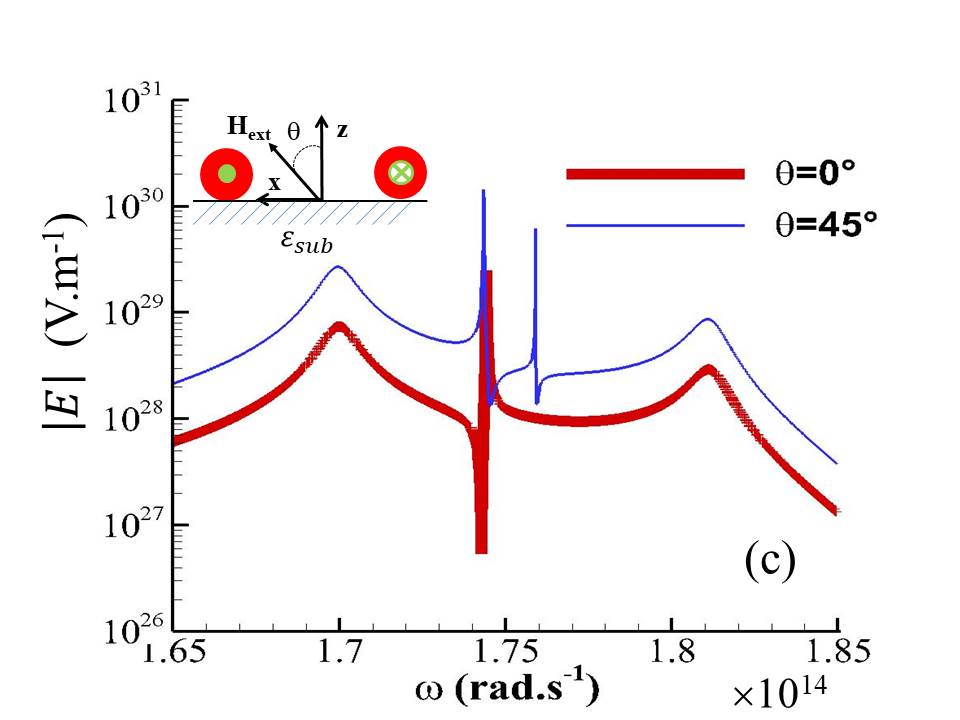}
\includegraphics[scale=0.3]{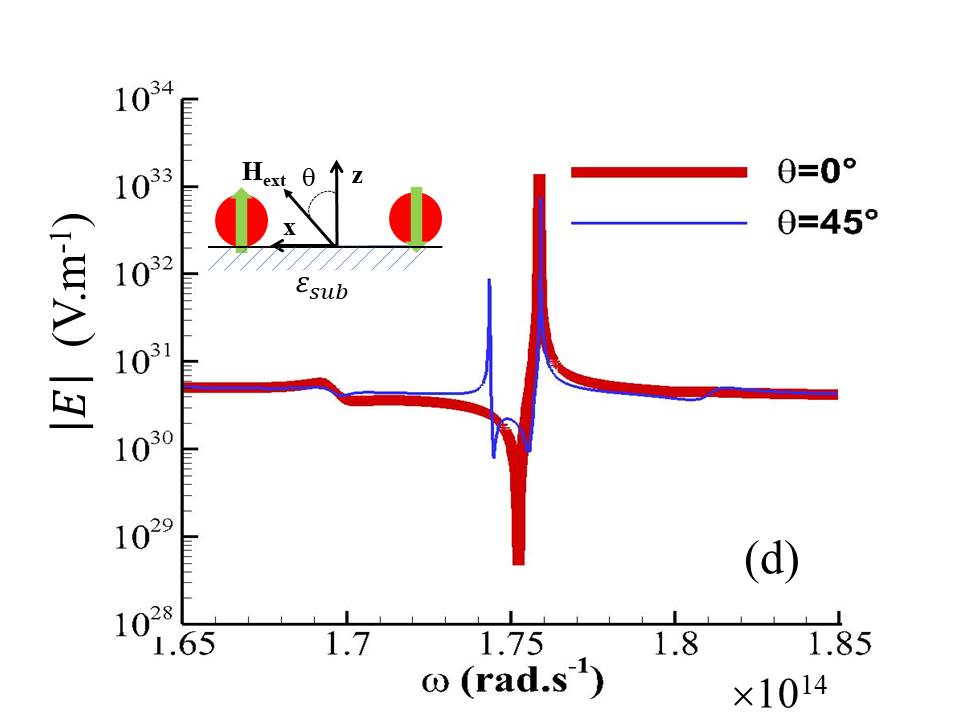}
\caption{(a)  LDOS in the same system as in Fig.\ref{ldos-dimer-vac} ($H_{ext}=5 T$) in presence of a substrate ($\varepsilon_{sub}=4$). (b)-(c) and (d)  Magnitude of electric field  radiated by two permanent (nomalized) dipoles oriented in opposite direction along the $\bold{x}$,  $\bold{y}$ and $\bold{z}$ axis, respectively.}
\label{ldos-dimer-sub}
\end{figure}

\begin{figure}
\includegraphics[scale=0.32]{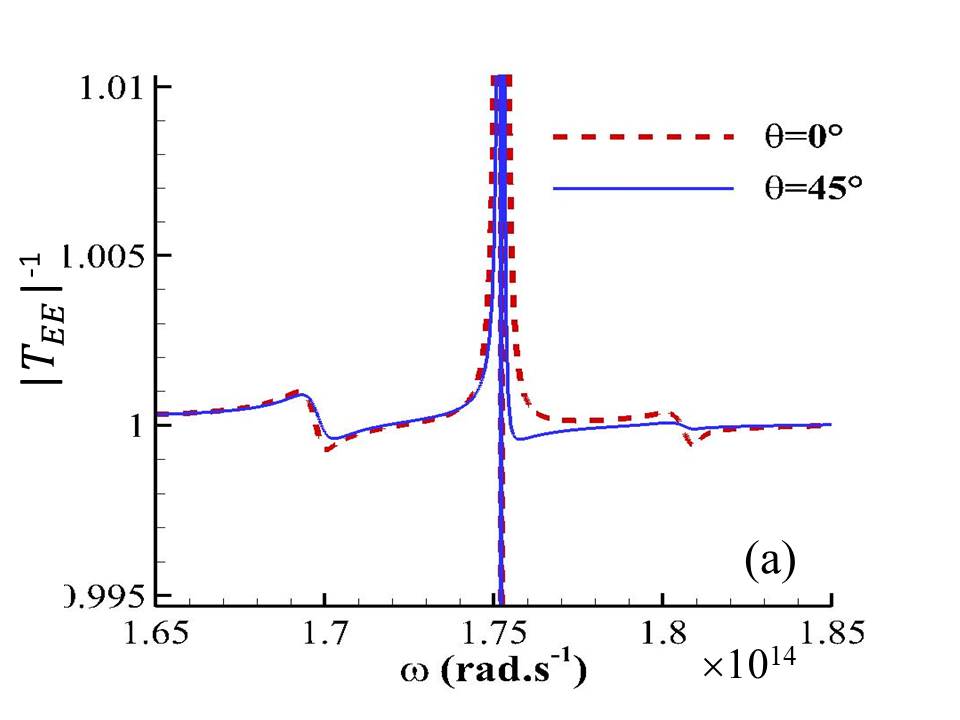}
\includegraphics[scale=0.32]{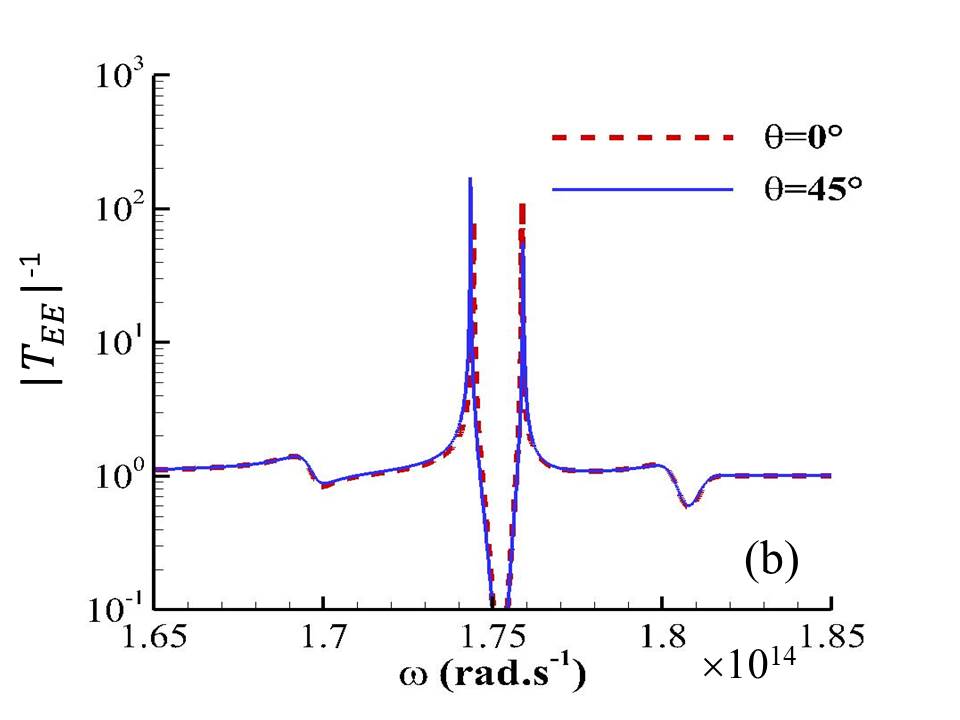}
\caption{ Resonance frequency for a dimer of InSb particles (radius $R=50\:nm$) separated by a distance $6=6R$ (a) in vacuum  and (b) on a transparent subtrate ($\varepsilon_{sub}=4$) under the action of an external magnetic field of magnitude $H_{ext}=5 T$ making an angle $\theta$ with the $\bold{z}$ axis. }
\label{reso-2InSb}
\end{figure}

To illustrate these changes we consider a simple dimer of InSb particles in vacuum and above substrate and we investigate its behavior under a change in the orentation of applied magnetic field. The LDOS at $z=3R$ between the two particles plotted in Fig.\ref{ldos-dimer-vac}-a shows the presence of a peak at the resonance frequency where $ \varepsilon_{3}=-2$ when the magnetic field is tilted with respect to the $\bold{z}$ axis while the LDOS remains elsewhere invariant with this rotation. To analyze the origin of this singular behavior we calculate the electromagnetic field radiated by permanent dipoles in given orientations under the action of external magnetic field. The  corresponding electric ans magnetic fields are calculated from expressions (\ref{Eq:field_fluc}) and (\ref{Eq:mag_fluc}) using (unitary) permanent dipoles rather than fluctuating dipoles.
The results are plotted in Fig. \ref{ldos-dimer-vac}-b and \ref{ldos-dimer-vac}-c when the dimer is in vacuum and the dipoles are oriented along the dimer axis. This calculation shows that both the electric and the magnetic fields are enhanced in the central region where we observe a drastic enhancement of the LDOS under a tilting of applied magnetic field. It is worthwhile to note that although the magnetic field present a resonance peak its magnitude is insufficient to contribute to the LDOS in the infrared, the latter being mainly driven by the electric field. Also, we observed  (not shown here)  that the fields radiated by the permanent dipoles oriented in all other directions in the canonical basis present similar features but these fields are less prominent and therefore do not contribute significantly to the LDOS.
In presence of a substrate (see Fig.\ref{ldos-dimer-sub}-a) the LDOS is overall enhanced and the central peak  ( Fig.\ref{ldos-dimer-sub}-a) hybridizes into two peaks at lower and higher frequency when $H_{ext}$ is tilted with respect to the normal of the surface while the peak at low frequency is almost imperceptible under normal incidence. Once again this behavior can be linked to the field radiated by the permanents dipoles oriented in all possible directions. These fields are plotted in Figs.\ref{ldos-dimer-sub}-b,c and d and they reveil that the peak at high frequency is mainly related to the permanent dipole oriented in the direction normal the surface (Fig.\ref{ldos-dimer-sub}-d). while the peak at low frequency results from the dipoles oriented both in the $\bold{z}$ and $\bold{x}$ directions (see Figs.\ref{ldos-dimer-sub}-b and \ref{ldos-dimer-sub}-d).
As shown in Figs.\ref{reso-2InSb} these peaks correspond to the resonance frequencies of the system which are solution of the transcendental equation
\begin{equation}
\mid\mathds{T}_{EE}(\omega)\mid^{-1}=0.\label{Eq:resonance}
\end{equation}
The comparison of resonances spectrum without substrate (Figs.\ref{reso-2InSb}-a) to the one  in presence of a substrate (Fig.\ref{reso-2InSb}-b) shows that, on one hand, the substrate is responsible for a hybridization of the central resonances into two resonances at higher and lower frequency and  is also responsible for an enhancement of resonance peaks in the LDOS as in the case of a single particle (dipole). It is worthwhile to note that a change in the orientation of external magnetic field mainly impacts the magnitude of resonance peaks and not the resonance frequencies themselve.

\subsubsection*{C. Triangular network of particles}

\begin{figure}
\includegraphics[scale=0.25]{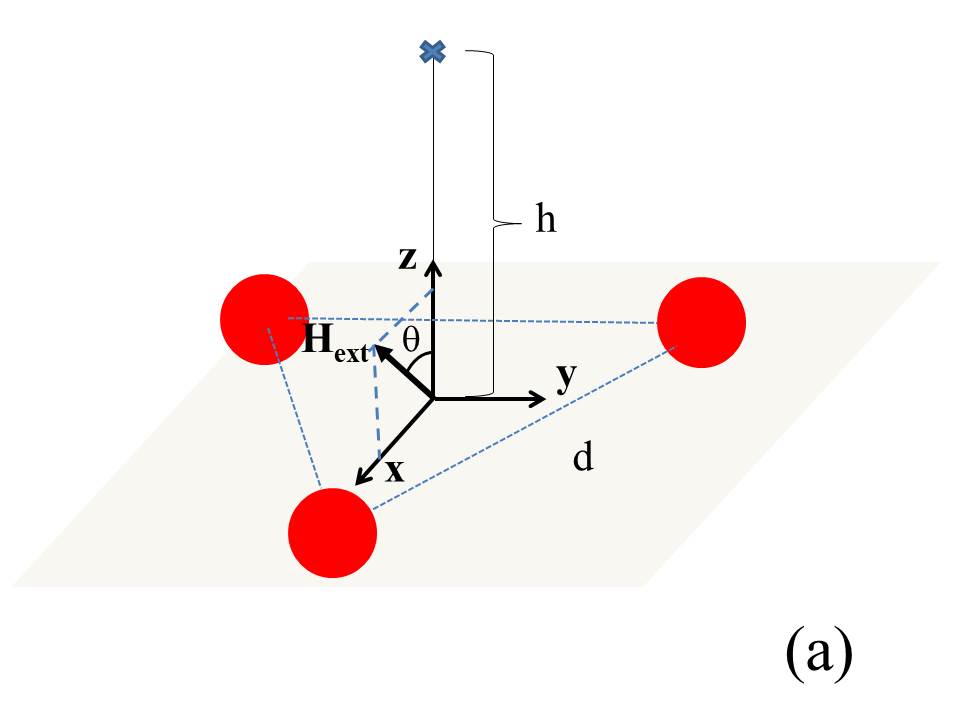}
\includegraphics[scale=0.3]{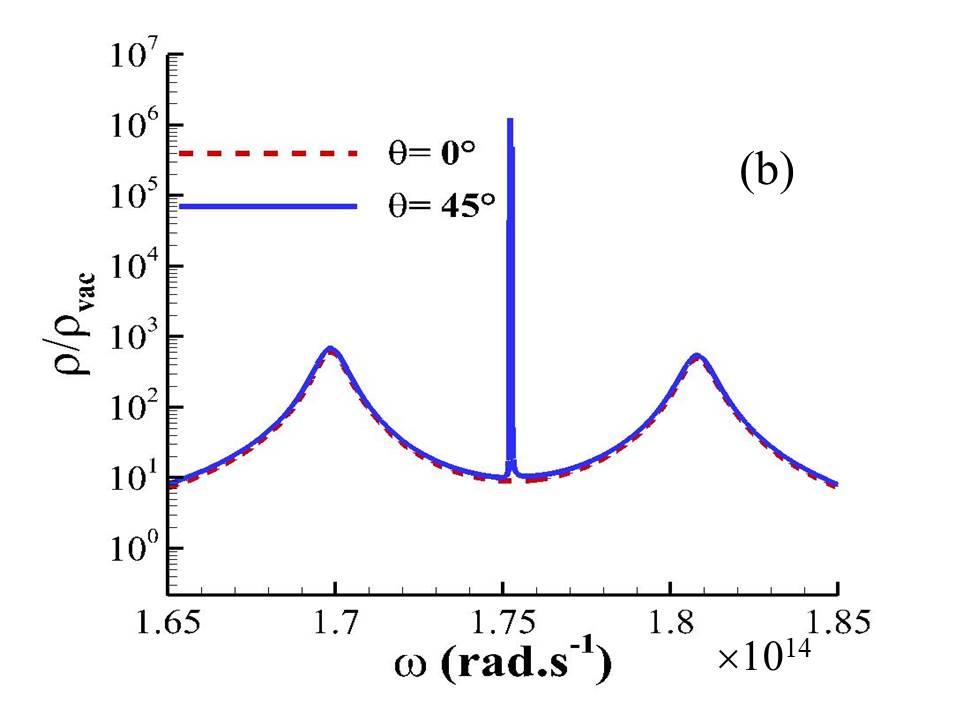}
\caption{ LDOS of electromagnetic field above  a regular network of InSb particles  under the action of an external magnetic field $H_{ext}$ making an angle $\theta$ with the $\bold{z}$ axis in the ($\bold{x}$,$\bold{z}$) plane. (a)  Particles network, regularly arranged, in vacuum, on an equilateral triangle inscribed inside a circle of radius $3R$ (i.e. triangle side of length $d=9R/\sqrt{3}$). (b)   LDOS at  $h=3R$ above the center of triangular network for two different orientation of magnetic field. The magnitude of magnetic field is $H_{ext}=5 T$.}
\label{triangle-vac}
\end{figure}

To finish we consider a regular triangular network of InSb particles (see Fig.\ref{triangle-vac}-a) both in vacuum or on a transparent substrate ($\varepsilon_{sub}=4$) and we show that the LDOS above this system can be efficiently tuned  (see Fig. Fig.\ref{triangle-vac}-b)  simply by changing the orientation or the magnitude of $\bold{H_{ext}}$. 
\begin{figure}
\includegraphics[scale=0.3]{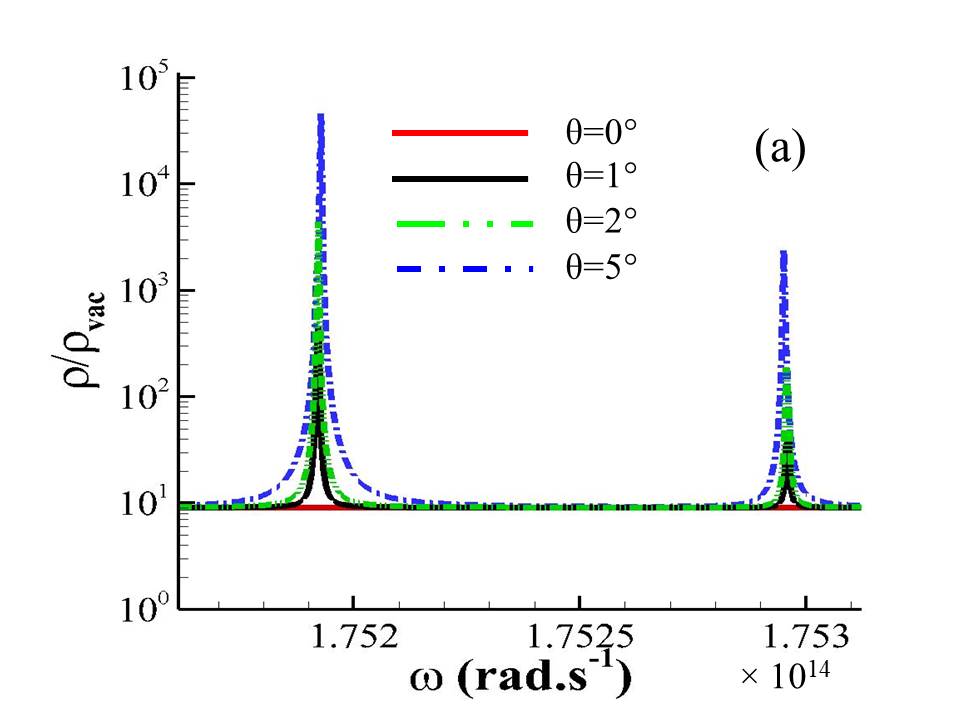}
\includegraphics[scale=0.3]{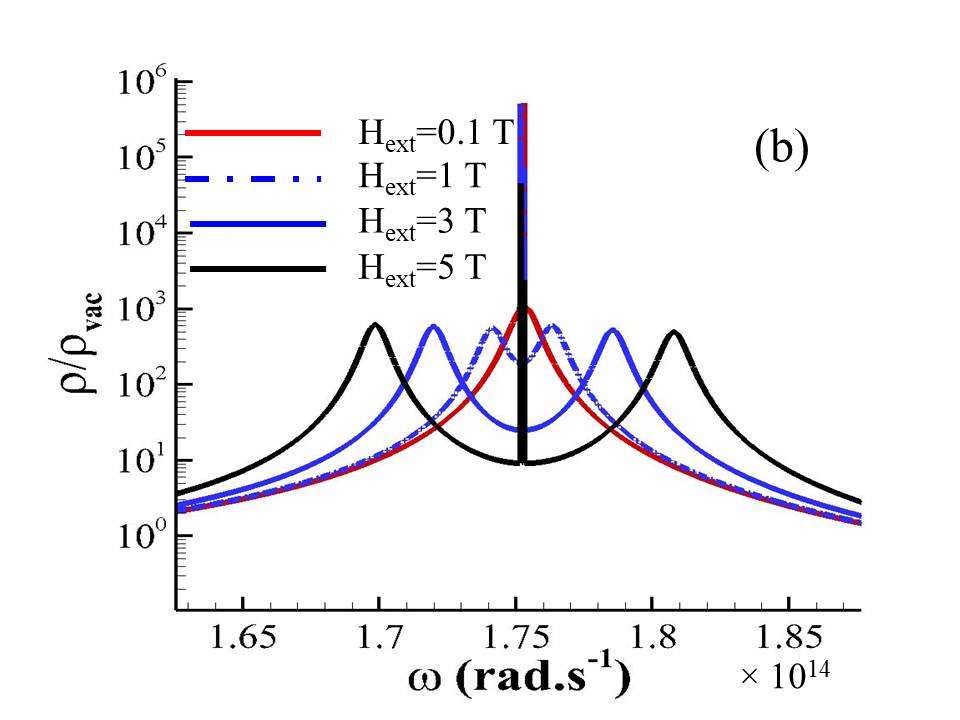}
\includegraphics[scale=0.3]{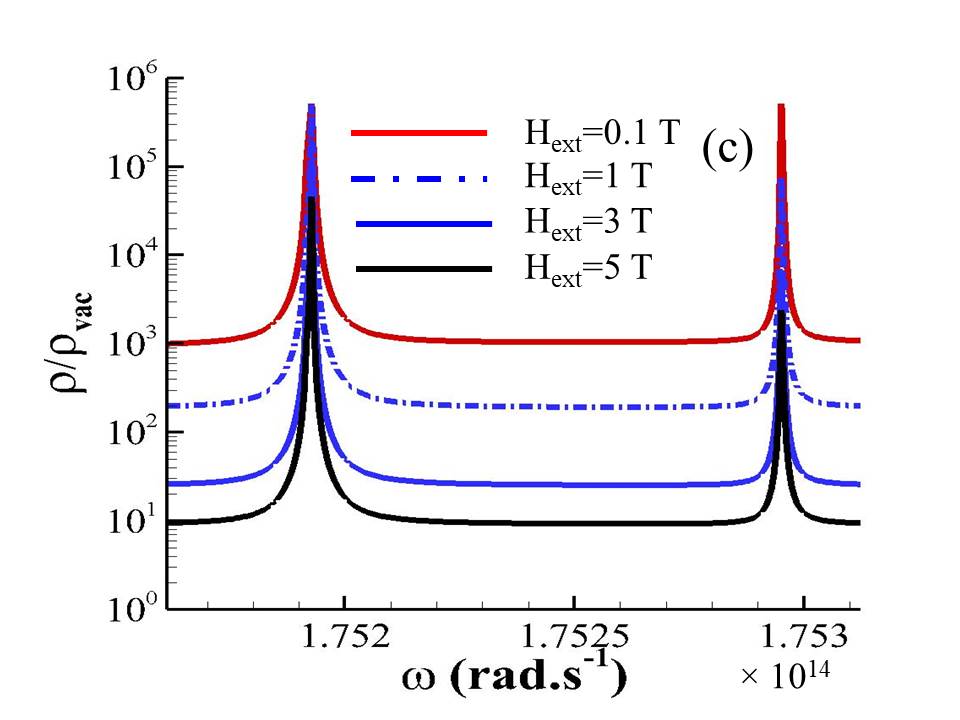}
\caption{ LDOS of electromagnetic field  at  $z=3R$ in vacuum above the same triangular network as in Fig.5 under the action of an external magnetic field $H_{ext}$ (a) vs its orientation angle $\theta$. The magnitude of magnetic field is $H_{ext}=5 T$. (b) LDOS vs the magnitude of $H_{ext}$ when $\theta=5^0$. (c) Zoom of Fig. 6-b in the central spectral region.}
\label{ldos_triangle_vs_param}
\end{figure}
\begin{figure}
\includegraphics[scale=0.3]{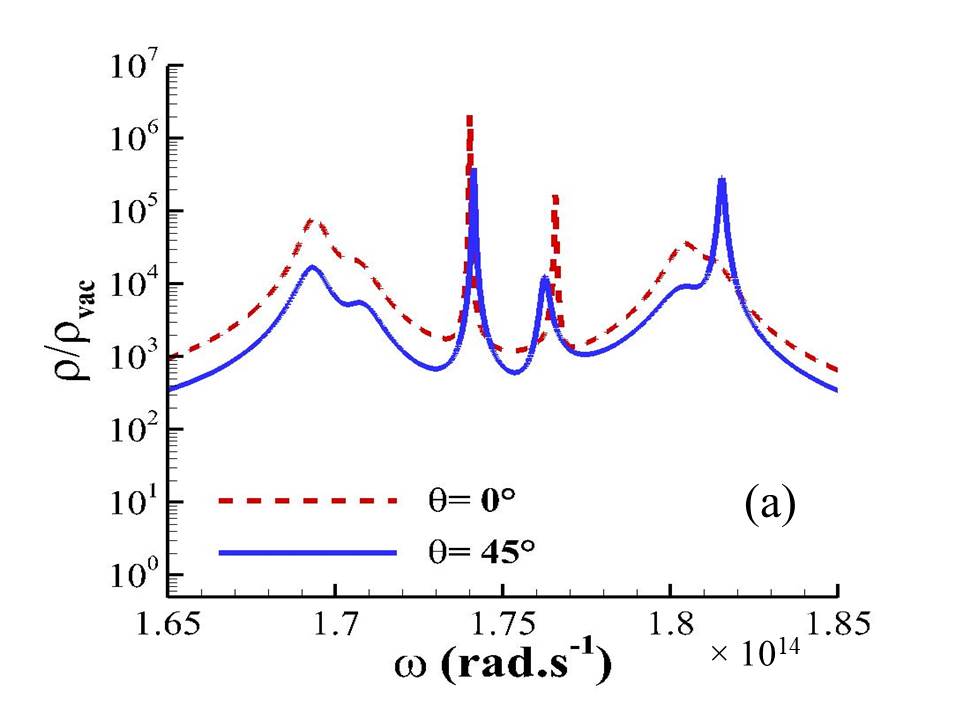}
\includegraphics[scale=0.3]{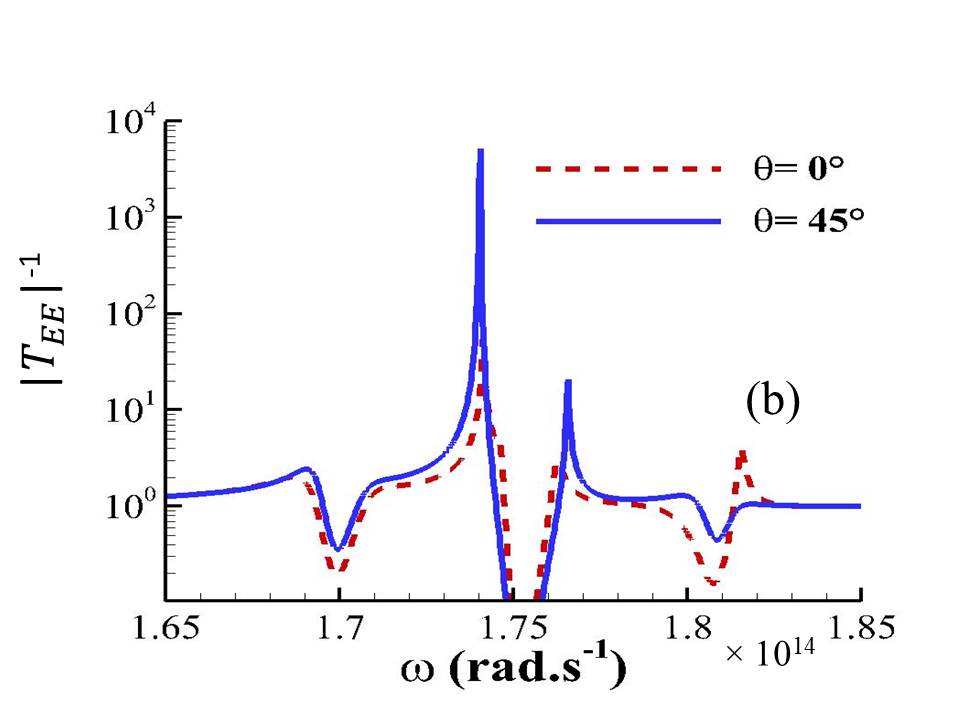}
\caption{ (a) LDOS of electromagnetic field above the same  network of InSb particles  as in Fig.\ref{ldos_triangle_vs_param}  but the network is placed on a transparent substrate ($\varepsilon_{sub}=16$). The magnitude of magnetic field is $H_{ext}=5 T$.(b) Resonance frequencies of network.}
\label{triangle-sub}
\end{figure}
\begin{figure}
\includegraphics[scale=0.3]{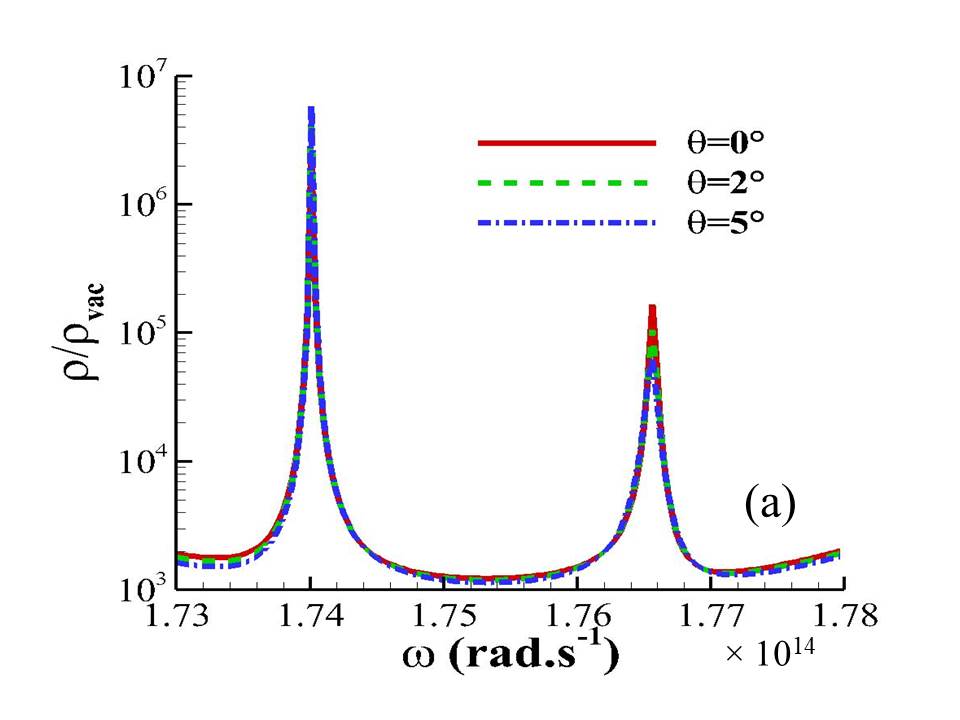}
\includegraphics[scale=0.3]{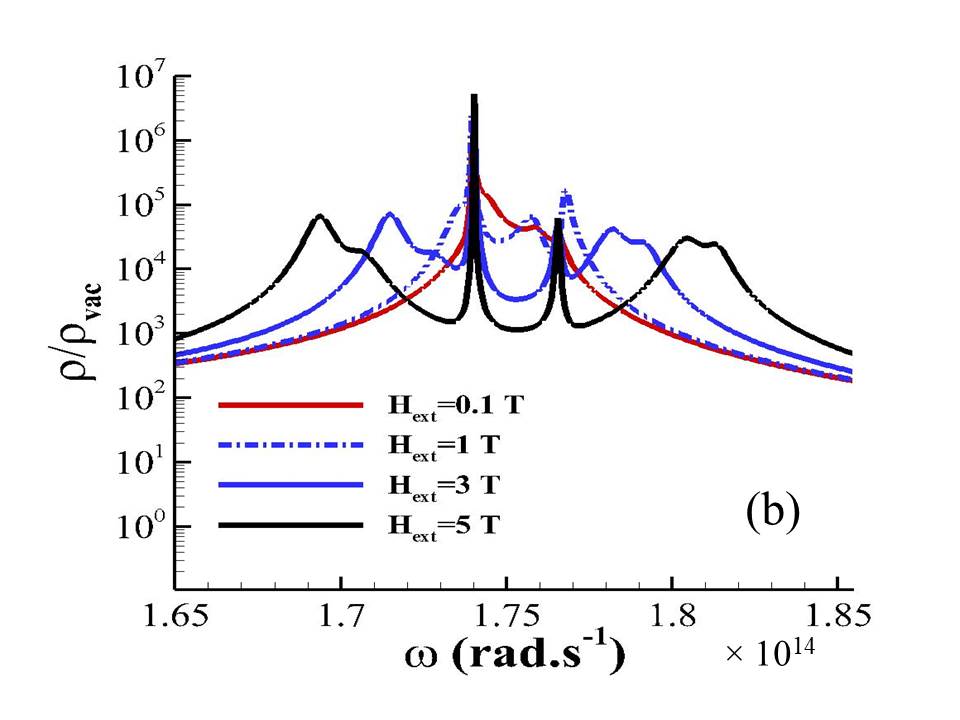}
\caption{ LDOS of electromagnetic field above a regular triangular network of InSb particles placed on a substrate as in Fig.\ref{triangle-sub}.  (a)   LDOS at  $h=3R$ above the center of network for different orientations of magnetic field when  $H_{ext}=5 T$. (b) LDOS vs the magnitude of $H_{ext}$ when $\theta=5^0$.}
\label{triangle-sub_param}
\end{figure}
When the set of particle is in vacuum ( Figs.\ref{triangle-vac}-b and \ref{ldos_triangle_vs_param}-a) we see that a tiny change in the orientation of external magnetic field allow for a drastic variation of the LDOS. Hence a rotation of $H_{ext}$  by only $\Delta\theta=1^0$  with respect to the normal of triangular network increases the value of the LDOS by almost two orders of magnitude. This increase rises up to more than four orders of magnitude with a rotation of  $\Delta\theta=5^0$ and even rises up to five orders of magnitude when $\theta=45^0$. In Figs. \ref{ldos_triangle_vs_param}-b and \ref{ldos_triangle_vs_param}-c, we see also that, due to the change of the location of resonance frequencies of  particles with the magntidue  of  $H_{ext}$, the LDOS can be modified over a broad spectral range by tuning this magnitude. When the network is on a transparent substrate, we see in Fig.\ref{triangle-sub}-a that, as for a dimer, the interface enhances the LDOS over all the spectral range. More interesting we see that the LDOS can be modulated by a factor of almost three over a broad spectral range simply by tilting the applied magnetic field by $45$ degrees.  For the highest frequency of hybridized modes (Fig.\ref{triangle-sub}-b) this modulation can even be modulated by one order of magnitude with this angular variation and by a factor $5$ (Fig.\ref{triangle-sub_param}-a) with a tilting of $H_{ext}$ by only $5$ degrees. As for the case were the network was in vacuum, the LDOS can be modulated by several orders  of magnitude over a broad spactral range simply by changing the magnitude of applied magnetic field (Fig.\ref{triangle-sub_param}-b).

\subsection*{IV. CONCLUSION}
In conclusion, we have investigated the photonic density of states near magnteo-optical metamaterials in presence of an external magnetic field and demonstrated that it can be significantly modified at subswavelength scale by changing the spatial orientation or the magnitude of this control parameter. We have shown that the contribution of different resonant modes supported by the structure can be selected by an ad hoc tunning of external magnetic field. 
These results provide a pathway for an active control of the LDOS and it could find broad applications ranging from the control of the spontaneous emission of quantum dots or atoms  to the manipulation of molecules  placed in the close vicinity of these media. By combining non-convex  optimization techniques~\cite{Chao} and knowledge in the areas of many-body interactions  a computational design of non-reciprocal  metamaterials could be performed in order to  make an ad hoc sculpting of the LDOS.

%
%

\begin{acknowledgments}
This work was supported by the  French Agence Nationale de la Recherche (ANR), under grant ANR-21-CE30-0030 (NBODHEAT). 
\end{acknowledgments}

\end{document}